# Effects of Porous Media Properties and Flow Environment on Drug Release from Porous Implants

Pawan Kumar Pandey,[1] KVS Chaithanya,[2] and Prateek K. Jha[1,†]

[1]Department of Chemical Engineering, Indian Institute of Technology Roorkee, Roorkee, Uttarakhand - 247667, India
[2]Department of Chemical Engineering, Indian Institute of Technology Indore, Indore, Madhya Pradesh - 453552, India

**ABSTRACT**. Drug-Filled Porous Implants (DFPIs) are an innovative solution for delivering drugs in a controlled and sustained manner to target sites. To optimize their performance across various physiological conditions, it is essential to understand how fluid flow and porous media properties influence the drug release process. In this work, we numerically investigate a wide range of flow conditions and their effects on drug release from DFPI. The DFPI is modeled as a homogeneous, saturated porous medium, with flow through the porous structure modeled using the Forchheimer-extended Darcy law. Drug diffusion within the DFPI and its transport through the surrounding channel are simulated using a diluted species transport approach. The results reveal the impact of flow conditions and porous media characteristics on the drug release profile of the implant and drug availability within the channel. The variations in drug release behavior are analyzed by modeling the release as an apparent first-order process with a time-dependent rate constant. Notably, the results highlight specific conditions under which the rate constant increases during the later stages of drug release from the DFPI, particularly at high Reynolds numbers, while also ensuring a prolonged operational time period of the implant. These findings suggest the potential for developing intelligent DFPI designs capable of delivering drugs in a manner more attuned to the specific needs of the application.

## I. INTRODUCTION.

Drug-Filled Porous Implants (DFPIs) are in widespread use in biomedical applications [1]. Several features of DFPIs like biocompatibility [2], porous structure [3] and the existence of multiple length scales [4] make them ideal drug reservoirs for controlled release. Specifically, their stimuli-responsive nature [5] helps achieve better control over the drug release kinetics [6]. For instance, modulating the drug release rate in response to external triggers is possible by incorporating stimuli-responsive elements into the DFPI matrix, which responds to temperature [7], pH [8,9], light [10], or specific biomolecules [11]. This helps maintain the drug levels in the target regions within the desired therapeutic window and enhances treatment efficacy [12]. Moreover, the polymer network of the DFPI, and consequently, its chemical and mechanical properties can be tuned to facilitate the shield mechanisms necessary for targeted drug delivery [13]. On the other hand, the porous structure of DFPIs can be easily tuned by controlling the cross-linking density and hydrophilicity (swellability), which enables control over the loading and release kinetics of the drug [14]. Therefore, the release kinetics of the drug from DFPI is sensitive to both the physical and chemical features of the porous media and its environment.

DFPI finds application in wide variety of biomedical applications - utilizing inherent or induced flow conditions to achieve targeted and controlled therapeutic release. In vascular systems, porous media embedded with therapeutic agents can leverage blood flow to facilitate sustained drug release, enhancing treatments for conditions such as cancer, hypertension, or atherosclerosis [15]. Similarly, in pulmonary systems, porous scaffolds or inhalable microspheres can release drugs under the influence of airflow, improving therapies for respiratory disorders like asthma or chronic obstructive pulmonary disease [16]. In the gastrointestinal tract, peristaltic movements and natural motility can aid in the diffusion of drugs from oral or injectable porous matrices, ensuring localized treatment for ulcers or infections [17]. In the eye, porous intraocular implants are used for sustained drug delivery over months in presence of aqueous humor flow [18]. By integrating the design of porous materials with an understanding of specific organ flow dynamics, it is possible to optimize DFPI properties for desired site-specific, prolonged therapeutic effects.

Drug release kinetics from the DFPI mainly depends on two length scales [19,20]: (i) the macroscopic length scale associated with the implant size and (ii) length scale of porous matrix. The macroscopic length scale determines the distance the drug molecule must travel to reach the surface whereas the microscopic length scale controls the rate of diffusion within the DFPI. The release kinetics depends on how the size of the drug molecule compares with these two length scales. Notably, both these length scales are sensitive to external stimuli, providing control over the release kinetics of the encapsulated drug. Thus, understanding the interplay of these length scales is crucial for optimizing the design and development of DFPI-based drug delivery systems, to achieve controlled and sustained release profiles that align with therapeutic requirements.

† Corresponding author: prateek.jha@ch.iitr.ac.in

The drug release kinetics are sensitive to several factors, namely: (i) the porosity of DFPI, (ii) the permeability of DFPI, (iii) the fluid flow through and around DFPI, and (iv) the effective diffusivity of the drug within DFPI. We investigate the effect of these parameters on the drug release kinetics using a two-dimensional model of drug release from DFPI placed in a channel with flow, shown in Figure 1. The range of parameters in current investigation (see Table 1) are selected to mimic the flow and implant properties of ocular and cardiovascular systems [18,21–25].

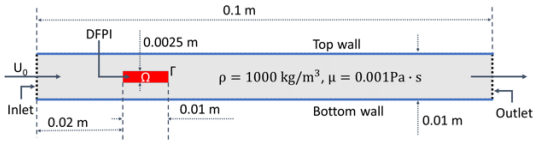

Figure 1: Schematic of a channel containing a drug filled porous implant (DFPI). The diagram shows dimensions of the channel and DFPI (represented by volume $\Omega$ and surface $\Gamma$), boundary labels, fluid's density ($\rho$) and viscosity ($\mu$), and direction of inflow and outflow. $U_0$ is the inlet velocity.

TABLE I. Model parameters and their values for different studied cases

| Parameter | Numerical values |
|---|---|
| Reynolds number (Channel), Re | 0.01, 0.1, 1, 10, 100 |
| DFPI Permeability (m$^2$), K | $10^{-9}$, $10^{-12}$ |
| DFPI Porosity, $\epsilon$ | 0.1, 0.3, 0.5 |
| Drug Diffusivity in clear media (m$^2$/s), $D_{clear}$ | $10^{-9}$ |
| Effective drug diffusivity in DFPI (m$^2$/s), $D_{eff}$ | $D_{clear} \cdot \epsilon^{\frac{4}{3}}, 10^{-13}$ |

## II. METHODS

The schematic of the DFPI placed in a channel is shown in Figure 1. The DFPI is positioned closer to the inflow boundary than the outlet to ensure that sufficient domain length is remaining downstream to allow for the flow to become fully developed again and minimize the axial variation of concentration.

Given the drug concentrations loaded in the DFPI for the cases analyzed in this study, the drug concentration in the surrounding solvent is considered dilute. Further, the DFPI is assumed to be always saturated; the effective drug diffusivity in Table 1 includes the dissolution step prior to diffusion. The drug release from the DFPI into the surrounding fluid is modeled using the Transport of Diluted Species (tds) and Laminar Flow (spf) modules in COMSOL Multiphysics 6.2.

The drug transport is governed by the following equation:

$$\frac{\partial c}{\partial t} + \mathbf{u} \cdot \nabla c = \nabla \cdot (D \nabla c) \qquad (1)$$

where c is the drug concentration (mol/m$^3$), D is the diffusivity of the drug, which has different values inside (D = $D_{eff}$) and outside (D = $D_{clear}$) the DFPI, and $\mathbf{u}$ is the fluid velocity. The first term on the left-hand side represents the accumulation rate of the drug, the second term accounts for convective transport, and the right-hand side represents diffusion.

To compute the effective diffusivity ($D_{eff}$) within the porous medium, the Bruggeman model is employed [26], which relates $D_{eff}$ to the clear medium diffusivity ($D_{clear}$) via the porosity ($\epsilon$) and tortuosity factor ($\tau$):

$$D_{eff} = D_{clear} \frac{\epsilon}{\tau} \qquad (2)$$

Here, the tortuosity factor can be estimated using multiple models, the model used in present work is as follows [26]:

$$\tau = \epsilon^{-1/3} \qquad (3)$$

The exponent in Equation (3) may vary depending on the pore structure of the porous medium or the DFPI. In most of the simulated cases, we have taken $-1/3$ as one of the possible values. That is, the effective diffusivity of drug within the DFPI is taken as

$$D_{eff} = D_{clear}\epsilon^{4/3} \quad (4)$$

However, in some of the cases discussed towards the end of the paper, we have assumed a constant value, $D_{eff} = 10^{-13}$ m$^2$/s, as also shown in Table 1. Convective transport of the drug requires flow field information, which is determined by solving the volume-averaged momentum conservation equation:

$$\frac{\rho}{\epsilon}\left(\frac{\partial \mathbf{u}}{\partial t} + (\mathbf{u} \cdot \nabla)\frac{\mathbf{u}}{\epsilon}\right) = -\nabla p + \nabla \cdot \left[\frac{1}{\epsilon}\{\mu(\nabla \mathbf{u} + (\nabla \mathbf{u})^T)\}\right] - \frac{\mu}{K}\mathbf{u} + \mathbf{F} \quad (5)$$

The additional drag force due to inertial effects, modeled by the Forchheimer term (**F**), is expressed as:

$$\mathbf{F} = -\beta\rho|\mathbf{u}|\mathbf{u} \quad (6)$$

where the Forchheimer coefficient (β) is calculated as:

$$\beta = \frac{c_F}{\sqrt{K}} \quad (7)$$

In these equations, ρ and μ represent the fluid density and dynamic viscosity, respectively, p is the pressure, and K is the permeability of the porous medium. The Forchheimer parameter ($c_F$) is set to 0.55 in this study. The flow is assumed to remain in the laminar regime, and thus, turbulence effects are not included in the model. The fluid is also considered incompressible, providing the continuity equation:

$$\nabla \cdot \mathbf{u} = 0 \quad (8)$$

At the left boundary (inlet) of the channel, a steady inlet flow condition with zero drug concentration is imposed:

$$\mathbf{u} = (U_0, 0) \quad (9)$$

$$c = 0 \quad (10)$$

At the outlet boundary (right side of the flow channel), the following conditions are applied:

$$p = 0 \quad (11)$$

$$\mathbf{n} \cdot (D\nabla c) = 0 \quad (12)$$

The top and bottom walls of the channel are assigned no-slip and no-flux boundary conditions for velocity and concentration, respectively.

The velocity (**u**) and pressure (p) are initialized to zero everywhere in the domain. Initial drug concentration is taken as zero outside the DFPI. However, inside the DFPI, initial concentration is taken as $1/\epsilon$. This initialization ensures consistency in the initial drug loading across different porosity (ϵ) values, facilitating a fair comparison of drug release profiles under varying porosity conditions.

We have validated the concentration transport model by comparing the bioavailability time period of drugs at target ocular tissue by comparing our results with values reported in the literature [24]. Details of this validation is reported in one of our previous works [18]. For brevity, grid independence and used mesh details are provided in the supplementary section. Figure S1 shows the related plots and the mesh distribution around the DFPI.

### III. RESULTS AND DISCUSSION

In this study, we explore the influence of porosity (ϵ), permeability (K), effective drug diffusivity ($D_{eff}$) within the DFPI, and the channel Reynolds number (Re) on the drug release kinetics.

#### A. Flow structures

Figure 2 illustrates the impact of Reynolds number on downstream flow structures. At Re=100, distinct recirculation regions are observed immediately downstream of the DFPI as shown in Fig. 2(b), whereas for Re=1 (Fig. 2(a)), Re=0.1, and Re=0.01, the streamline structures remain nearly identical, with negligible differences (hence, only results for Re=1 and Re=100 are shown). In this range of Reynolds numbers, the flow remains steady, with no signs of unsteady behavior or periodic flow instabilities downstream of the DFPI. However, a brief initial period of flow development occurs, during which the conditions are transient.

Similarly, variations in permeability K, ranging from $10^{-12}$ m$^2$ to $10^{-9}$ m$^2$, do not significantly alter the streamline patterns. However, the Reynolds Number (Re) in channel and porous media conditions significantly influence the flow patterns within the DFPI, as illustrated schematically in Figure 3. The key differences are observed in the diverging and converging streamlines exiting from the right face (downstream) of the DFPI. Furthermore, outflow occurs through the top and bottom faces up to a certain distance from the inflow (left) face. The vertical dashed line in Figure 3 indicates transition from outflow patch to inflow patch along the top and bottom faces. The extent of the outflow region (indicated by span between left face and vertical dashed line in Figure 3) increases with a decrease in either the permeability or the Reynolds number. Notably, the effect of permeability is more pronounced compared to the Reynolds number in determining the outflow span.

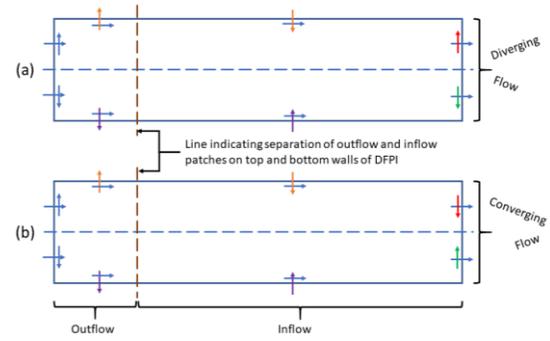

Figure 3: The incoming and outgoing flow patterns in DFPI. Two distinct patterns appear for different cases (see text): (a) streamlines diverge at the downstream face of DFPI (b) streamlines converge at the downstream face of DFPI

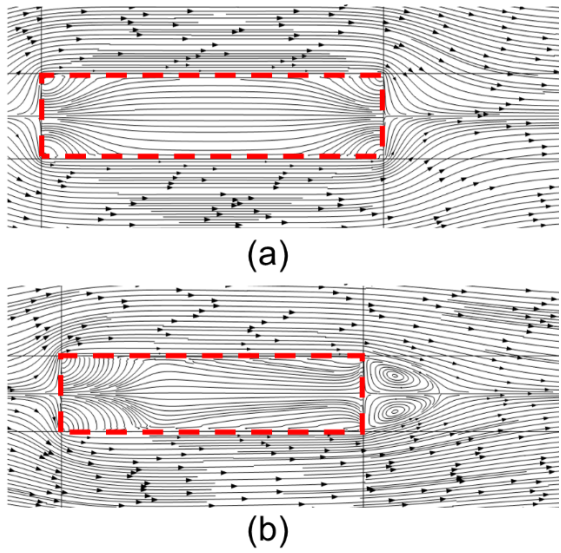

Figure 2: Flow structures around the DFPI for (a) Re = 1 and (b) Re = 100. Flow structures are visualized using streamlines for porous medium properties: $K = 10^{-12}$ m$^2$, $\epsilon = 0.1$, $D_{eff} = D_{clear} \cdot \epsilon^{4/3}$.

For low permeability cases ($K = 10^{-12}$ m$^2$), the flow patterns exhibit divergence at the downstream face for Reynolds numbers of 100 and 10, as shown in Figure 3(a). In contrast, for high permeability cases ($K = 10^{-9}$ m$^2$), diverging flow patterns are observed only for Re=100, as depicted in Figure 3(b). The diverging patterns on the downstream face indicate the presence or onset of downstream recirculation. The converging or diverging flow on the right face of DFPI results from the wake structure formed behind it. While lower Re flows exhibit no recirculation, higher Re cases (Re=100) show clear recirculation zones in the wake region, aligning with observations reported by Seol et al. [27]. Although the present work does not explicitly examine the effects of confinement walls, prior studies have highlighted their influence on flow behavior and thermal transport [28].

At the channel inlet, we set a uniform velocity profile, which gradually develops along the length of the channel. Figure 4(a) shows the x-component of velocity along the centerline of the channel. Since we varied the Reynolds number (Re) by changing the inflow velocity, it is difficult to compare the absolute velocity profiles directly. To make the data easier to interpret, we normalized the velocity by dividing it by the inflow velocity. Once the flow becomes fully developed, the normalized velocity magnitude reaches a level of approximately 1.5. We focused on one set of porous medium properties because changing the porosity do not significantly affect the velocity profile. Permeability did have a small impact on the normalized velocity inside the DFPI, but overall, the shape of the velocity profile stayed mostly the same. The results show that, except for cases with Re=100, all other cases achieve fully developed flow before encountering the porous DFPI. As expected, the velocity decreases noticeably as the flow enters the porous DFPI. Figure 4(a) shows that after passing through the DFPI, the flow regains its structure and becomes fully developed again for Re=0.01 to Re=10. However, for Re=100, it takes a little longer—roughly an additional half-channel length—for the flow to fully settle, which is specific to the chosen geometric dimensions. Downstream of the DFPI, the velocity curve shows a small "inverted V" shape, which indicates the formation of a recirculation zone in the wake. Near the outlet of the channel, slight oscillations in the velocity can be seen, particularly in the range $0.095 \leq x \leq 0.1$ (Figure 4(a)). This confirms that the domain length is sufficient to keep the effects of the outflow boundary confined to the outlet region. Figure 4(b) presents the velocity distribution within the DFPI, normalized by the respective inflow velocities. For cases with higher Re, the normalized velocity is noticeably higher near the upstream end of the DFPI and lower near the downstream end. This distribution can be attributed to the increased capacity of high-inertia flows to penetrate the porous medium at the upstream interface. Conversely, the reduced normalized velocity at the downstream end for higher Re is a consequence of the enhanced resistance introduced by inertial effects within the porous structure [29].

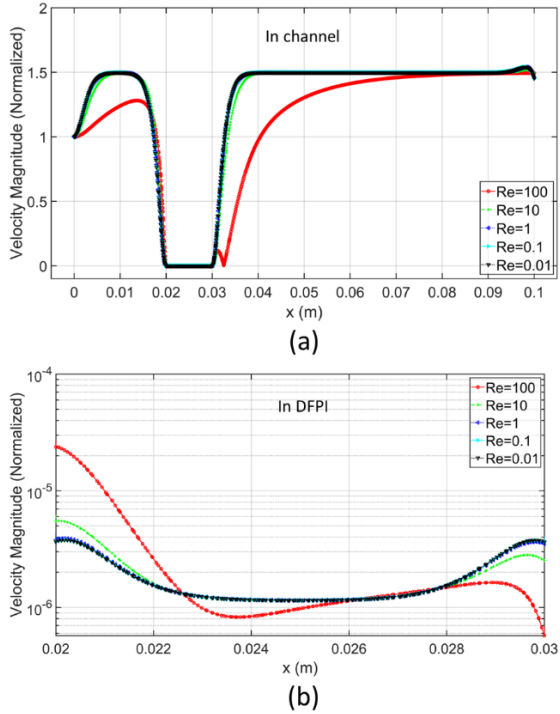

Figure 4: x-component of velocity vs. x, along the horizontal centreline passing through channel, in (a) the channel, and (b) zoomed-in view inside the DFPI. Results are compared for different Reynolds number (Re) for porous medium properties: $K = 10^{-12}$ m$^2$, $\epsilon = 0.1$, $D_{eff} = D_{clear} \cdot \epsilon^{4/3}$ m$^2/s$.

### B. Drug Release Behavior

Figure 5 shows how the drug concentration, normalized with initial concentration, inside the DFPI decreases over time for different cases. To capture the wide range of variations, the data is plotted on a semi-log scale. In the following, we have defined the "depletion time" as time after which the drug concentration in DFPI falls below 0.01% of initial level. As shown in Figures 5(a) and 5(b), increasing the Reynolds number or porosity tends to shorten the depletion time. The rate of drug release from the DFPI is governed by the concentration gradient between the DFPI and the channel, which depends on drug clearance from the channel. Moreover, because of varying release rate, the depletion time also depends on the chosen cut-off level (0.01% in our case). Therefore, the observed trends may not be generalized to all choices of porosity and Reynolds number.

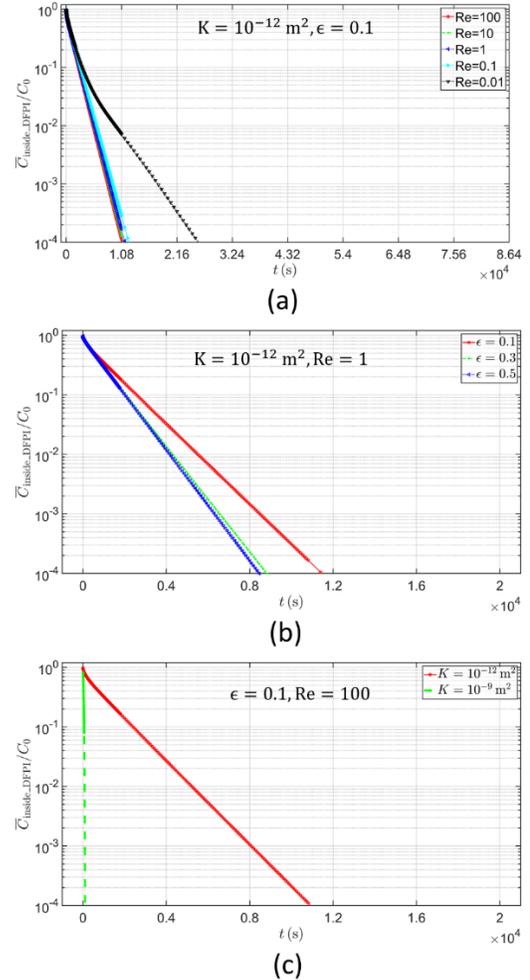

Figure 5: Average drug concentration (normalized) inside the DFPI with time for (a) different Reynolds number (Re) with $K = 10^{-12}$ m$^2$, $\epsilon = 0.1$, $D_{eff} = D_{clear} \cdot \epsilon^{4/3}$. (b) different porosity ($\epsilon$) with $K = 10^{-12}$ m$^2$, Re = 1, $D_{eff} = D_{clear} \cdot \epsilon^{4/3}$; (c) different permeability (K) with $\epsilon = 0.1$, Re = 100, $D_{eff} = D_{clear} \cdot \epsilon^{4/3}$.

In cases with low permeability shown by the red line in Figure 5(c), there seems to be a lower limit ($t \approx 1.08 \times 10^4$ s, also see Figure 5(a)) for the depletion time. However, for high-permeability cases, as shown by the green line in Figure 5(c), this lower limit is no longer observed. This is because, in high-permeability scenarios, convective effects dominate, and drug depletion resembles a sweeping action driven by the flow (will be discussed later). Consequently, the influence of other parameters, apart from inertial effects, becomes negligible. Interestingly, the depletion trend of concentration inside the DFPI shows unexpected curvatures (Figures 5(a)-(c)). The presence of these curvatures suggests a deviation from ideal first-order drug release, which may have implications for practical applications.

In practical applications like drug-coated stents and ocular implants, it is essential to understand how drug concentration varies at the target site. In this study, we calculated the average drug concentration outside the DFPI but within the whole rectangular channel, representing the diseased area. Figure 6 shows how this average concentration (normalized with initial DFPI concentration) changes over time. In all cases, the concentration rises quickly at first, then levels off into a plateau during the middle phase, and finally decreases in the later stages. The initial rise may be attributed to the influx of drug from the DFPI and the later decline may be attributed to the drug clearance from the channel outlet. The plateau is attained when the drug influx from the DFPI is approximately balanced by its subsequent convection outside DFPI. Such plateau is therefore not observed at higher Re, due to strong convection outside the DFPI. On the other hand, for lower Re where the drug clearance is slower, the plateau phase lasts longer, and the decline is delayed, as shown in Figure 6(a). Interestingly, while permeability has little impact on the average concentration (see Figures 6(c)), porosity plays a significant role, as shown in Figure 6(b). This is due to the faster drug release from the DFPI at higher porosity. However, this observation cannot be broadly generalized, as the overall outcome depends on the interplay between drug release from the DFPI, clearance from the channel, and the temporal lag between these processes.

Higher Reynolds number combined with higher permeability exhibits a dominant convective effect inside the DFPI. This necessitates examining the spatial distribution of the drug inside the DFPI. Figure 7 illustrates the drug concentration within the DFPI at time instants corresponding to 75%, 50%, and 25% of the drug remaining. For low permeability cases (Figure 7(a)), drug depletion occurs uniformly across all four faces of the DFPI, with a slight exception at the left face. In contrast, high permeability cases (Figure 7(b)) reveal markedly different depletion patterns for Reynolds numbers Re=1,10, and 100, exhibiting behavior akin to drug sweeping by the flow. Unlike the low permeability cases, the high permeability scenarios develop two distinct regions: one where the drug is fully depleted and another where the drug concentration remains same as its initial level. The depletion front contracts more rapidly in the flow direction compared to the transverse direction.

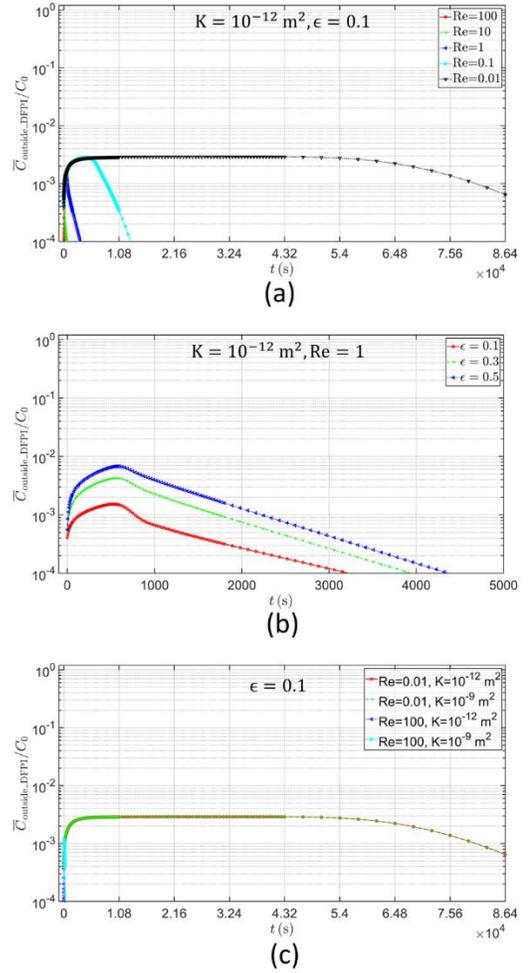

Figure 6: Average drug concentration (normalized) outside the DFPI with time for (a) different Reynolds number (Re) with $K = 10^{-12} \text{ m}^2, \epsilon = 0.1, D_{eff} = D_{clear} \cdot \epsilon^{4/3}$ ; (b) different porosity ($\epsilon$) with $K = 10^{-12} \text{ m}^2, Re = 1, D_{eff} = D_{clear} \cdot \epsilon^{4/3}$ ; (c) different Permeability (K) with $Re = 0.01, 100, \epsilon = 0.1, D_{eff} = D_{clear} \cdot \epsilon^{4/3}$ .

The influence of flow structures, as discussed in Figure 3, is evident when comparing the concentration profiles for high permeability cases at varying Reynolds numbers. Notably, the drug concentration contours exhibit diverging and converging shapes at the downstream (right) face of the DFPI, consistent with the diverging and converging streamlines observed for Re=100 and Re=10, respectively. At higher Reynolds numbers, the concentration front adopts a concave shape, with the degree of concavity diminishing as Re decreases, eventually transitioning to convex shapes for Re=0.1 and Re=0.01.

In the convection-dominated flow regime, characteristic of high permeability cases, the drug is transported in bulk along the length of the DFPI. However, this flow regime not only enhances the convective flux of the drug but also increases the diffusive fluxes from the top, bottom, and right faces due to sharp concentration gradients along these boundaries. This behavior contrasts with the left face, where diffusive fluxes remain minimal.

The flow structures within the DFPI have significant implications for drug release in high permeability cases. As the drug is swept along the length of the DFPI, the convective flux diminishes with the progressive depletion of the implant. Around the point where 75%–50% of the initial drug concentration remains (with exact values depending on case parameters), convective drug transport ceases from the top and bottom faces, becoming restricted to the right face. This transition occurs as the drug front surpasses the outflow span of the top and bottom walls, as depicted in Figure 3.

For Re=10 and Re=100, the drug front advances rapidly, while the drug concentration within the red regions of the colormap shows minimal reduction. In the Re=1 case, the front advances while simultaneously depleting the core region. For Re=0.1 and Re=0.01, the depletion dynamics closely resemble those observed in the low permeability cases.

Figure 8 shows the drug concentration distribution in the entire channel (including DFPI) at the time instant when the 75% of initially loaded drug is remaining in the DFPI. Since the results are shown based on the same amount of drug remaining in the implant – contour for different cases correspond to different time instants. Interestingly, higher concentration levels are observed outside the DFPI at higher Reynolds numbers (Re).

As noted earlier, drug release is not perfectly following the first order release. However, to assess the deviations, we calculate the varying rate constant (k) based on apparent first order release equation:

$$k = \frac{\ln(\bar{C}_{DFPI,t}) - \ln(\bar{C}_{DFPI,t+\Delta t})}{\Delta t} \qquad (13)$$

where $\bar{C}_{DFPI,t}$ and $\bar{C}_{DFPI,t+\Delta t}$ are average concentration (normalized with initial concentration) inside DFPI at two consecutive time instants in the interval of $\Delta t$.

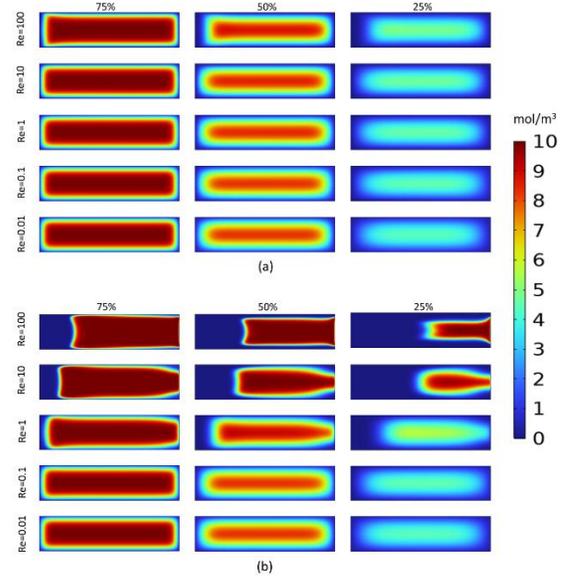

Figure 7: Drug concentration in DFPI at time instants when drug is 75%, 50%, and 25% remaining in DFPI. Results are shown for (a) $K = 10^{-12} \text{ m}^2, \epsilon = 0.1, D_{eff} = D_{clear} \cdot \epsilon^{4/3}$ and (b) $K = 10^{-9} \text{ m}^2, \epsilon = 0.1, D_{eff} = D_{clear} \cdot \epsilon^{4/3}$

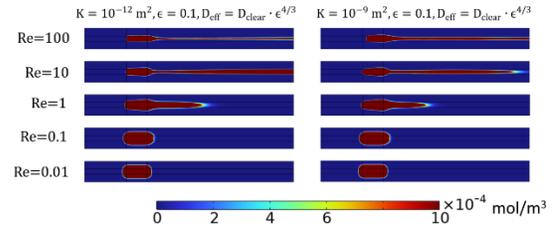

Figure 8: Drug concentration in whole domain at time instants when drug is 75% remaining in DFPI.

Figure 9 shows how k changes over time. As expected, k is highest at the start of the release process. For low-permeability cases (Figure 9(a)), k steadily decreases as the drug in the DFPI depletes. However, in high-permeability cases (Figure 9(b)), k initially drops but then begins to rise after a certain point, with the sharpest increases seen for higher Re. This behavior can be attributed to the relative contributions of convective and diffusive drug fluxes from the DFPI. Interestingly, during the early stages of drug release, k is nearly the same for all porosity cases for a given Reynolds number. But at later stages, porosity has a much more noticeable effect, as shown in Figure 9(c) for Re=1.

### C. Finding conditions with rising rate constant and slow drug depletion

The pattern of an initial decrease followed by an increase in k opens up exciting possibilities for designing DFPIs that deliver drugs in a controlled "on-and-off" manner, tailored to stay within a therapeutic window. This effect is most prominent at higher Reynolds numbers, such as Re=100, though it lasts for a shorter period compared to other cases (Figure 9(b)).

Comparison of Figures 9(a) and 9(b) suggests there is a noticeable transition between low- and high-permeability cases in the DFPI. The observed transition is likely caused by increased convection inside the DFPI while diffusion remains unchanged. To explore this intermediate stage — where the implant depletes more slowly but maintains a convection-to-diffusion ratio similar to the high-permeability case in Figure 9(b) — we ran additional simulations. In these cases, the permeability was set to $10^{-12}$ m$^2$, and the effective diffusivity of the drug in the DFPI was taken as low values of $10^{-13}$ m$^2$/s. The results for these cases are shown in Figures 10(a) and 10(b). The results shown in Figure 10(a) reveal that for higher Reynolds numbers, the rate constants increase over time, while the implants take significantly longer to deplete. Figure 10(b) highlights the low but sustained total flux of drug concentration coming out of DFPI surface (Γ) in contrast to the flux observed in the high-diffusivity cases with same permeability value, as shown in Figure 10(c).

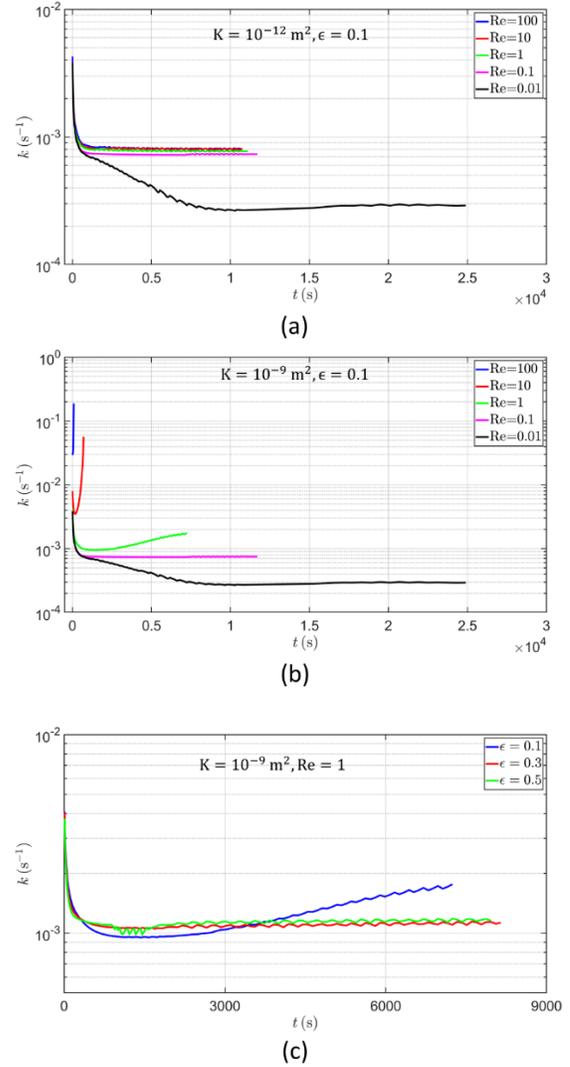

Figure 9: Variation of first order rate constant with time for (a) different Reynolds number (Re) with $K = 10^{-12}$ m$^2$, $\epsilon = 0.1$, $D_{eff} = D_{clear} \cdot \epsilon^{4/3}$; (b) different Reynolds number (Re) with $K = 10^{-9}$ m$^2$, $\epsilon = 0.1$, $D_{eff} = D_{clear} \cdot \epsilon^{4/3}$; (c) different porosity ($\epsilon$) with DFPI properties: $K = 10^{-9}$ m$^2$, Re $= 1$, $D_{eff} = D_{clear} \cdot \epsilon^{4/3}$.

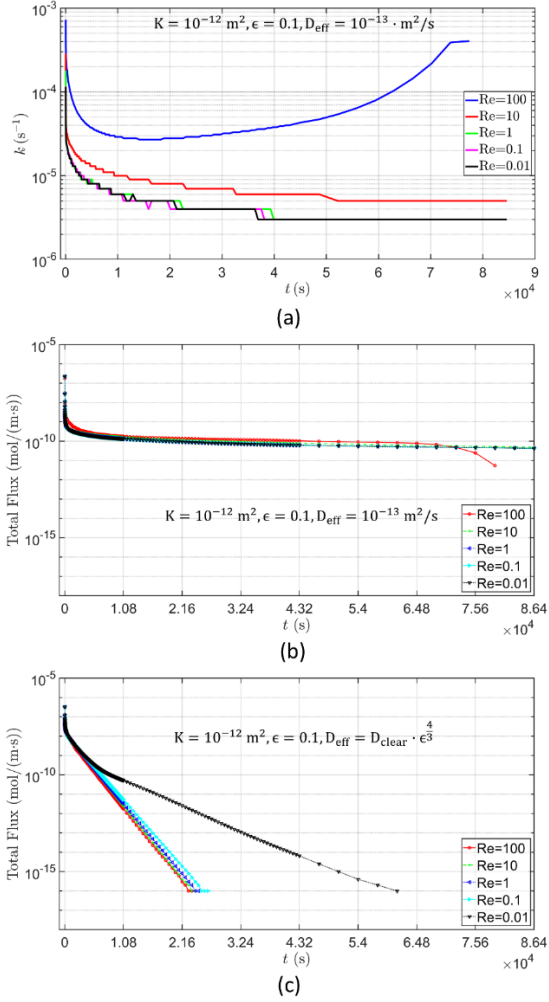

Figure 10: Variation of (a) rate constant (k) with time; (b) total flux with time for different Reynolds number (Re) with DFPI properties: $K = 10^{-12}$ m$^2$, $\epsilon = 0.1$, $D_{eff} = 10^{-13}$ m$^2/s$; and (c) total flux with time for different Reynolds number (Re) with DFPI properties: $K = 10^{-12}$ m$^2$, $\epsilon = 0.1$, $D_{eff} = D_{clear} \cdot \epsilon^{\frac{4}{3}}$.

Figure 11(a) demonstrates that a combination of lower permeability and lower effective diffusivity in the DFPI not only achieves an increasing rate constant with extended depletion time but also maintains a similar overall drug distribution within the implant, comparable to cases with higher permeability and effective diffusivity. This approach additionally results in higher drug concentration availability within the channel (outside of DFPI). When comparing the concentration contours outside the DFPI in Figure 11(b) with those in Figure 8, a distinct difference emerges: for cases with lower effective diffusivity (Figure 11(b)), higher drug concentrations are sustained at downstream locations, particularly at lower Reynolds numbers. This behavior contrasts with the trends observed in Figure 8. It is worth noting that, while all contour plots correspond to the point at which 75% of the drug remains in the implant, the time required to reach this stage varies significantly across the different cases.

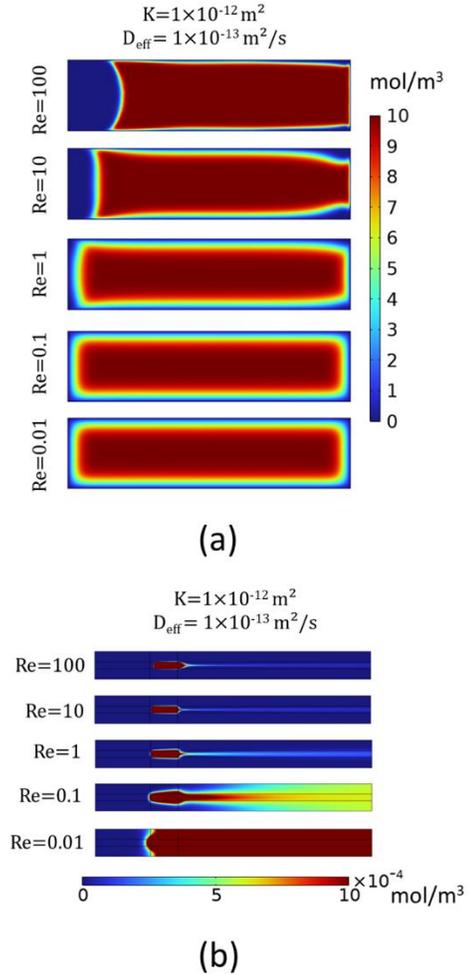

Figure 11: Drug concentration at time instants when drug is depleted to 75% level (a) in DFPI; (b) in channel for different Reynolds number (Re) with $K = 10^{-12}$ m$^2$, $\epsilon = 0.1$, $D_{eff} = 10^{-13}$ m$^2/s$;

The interesting behaviour of rate constant and drug depletion time period in different sets of porous medium properties of DFPI with Re=100, prompts us to focus on the drug depletion dynamics of following cases:

**Case 1 (low permeability + high effective diffusivity):**
$K = 10^{-12}$ m$^2$, $\epsilon = 0.1$, $D_{eff} = D_{clear} \cdot \epsilon^{\frac{4}{3}}$, Re = 100;

**Case 2 (high permeability + high effective diffusivity):**

$K = 10^{-9}\,\text{m}^2$, $\epsilon = 0.1$, $D_{\text{eff}} = D_{\text{clear}} \cdot \epsilon^{\frac{4}{3}}$, $Re = 100$;

**Case 3 (low permeability + low effective diffusivity):**

$K = 10^{-12}\,\text{m}^2$, $\epsilon = 0.1$, $D_{\text{eff}} = 10^{-13}\,\frac{\text{m}^2}{\text{s}}$, $Re = 100$;

The variation in drug depletion for these three cases— is shown by comparing the drug concentration along the DFPI centerline in Figure 12(a) for different time instants corresponding to drug remaining at the level of 90% to 10% at the interval of 10%. Interestingly, the drug distribution within the DFPI at various stages of depletion remains similar for the case 2 and case 3. In both cases, the transport of concentration exhibits a sweeping behavior. However, for the case 3, the concentration fronts are noticeably sharper compared to the case 2. This sharpness can be attributed to the lower effective diffusivity, which limits diffusion across the concentration fronts.

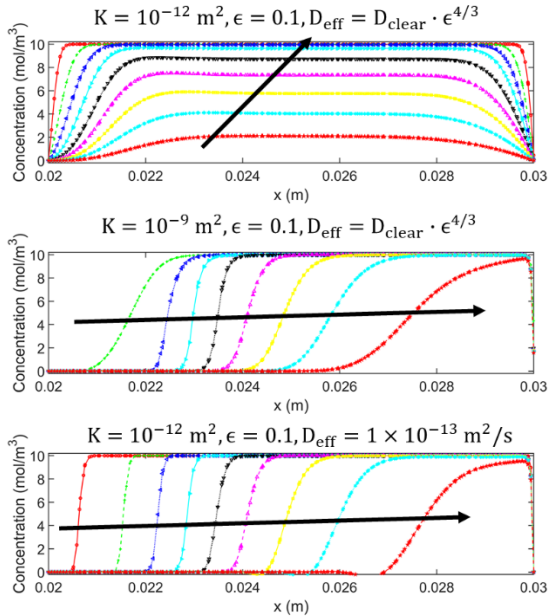

(a)

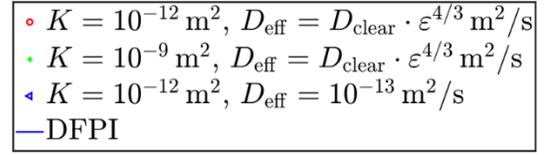

(b)

Figure 12 (a) Drug concentration in DFPI along the horizontal centreline is compared for cases with case 1 ($K = 10^{-12}\,\text{m}^2$, $\epsilon = 0.1$, $D_{\text{eff}} = D_{\text{clear}} \cdot \epsilon^{\frac{4}{3}}$), case 2 ($K = 10^{-9}\,\text{m}^2$, $\epsilon = 0.1$, $D_{\text{eff}} = D_{\text{clear}} \cdot \epsilon^{\frac{4}{3}}$), and case 3 ($K = 10^{-12}\,\text{m}^2$, $\epsilon = 0.1$, $D_{\text{eff}} = 10^{-13}\,\frac{\text{m}^2}{\text{s}}$); plots are shown for 9 different time instants starting at time when there was 90% drug left in DFPI to 10% drug left in DFPI at the interval of 10%; (b) Comparison of iso-contour lines of three different concentration levels for cases 1, 2 and 3 at time instants when 80% drug is remaining in DFPI.

In addition to influencing front sharpness, effective diffusivity also affects the concavity of the concentration fronts. Figure 12(b), which illustrates iso-contour lines for three concentration levels, highlights the reduction in front concavity with increasing effective diffusivity. This behavior underscores the role of effective diffusivity in modulating the spatial distribution of drug concentration during the depletion process.

The similarity in contour shapes between cases 2 and 3, as compared to case 1, can be attributed to the comparable convective-to-diffusive flux ratio, maintained by keeping $K/D_{eff}$ of the same order in cases 2 and 3. However, the extended drug depletion time in case 3—resembling case 1 but differing from case 2—is due to diffusion predominantly governing the depletion time estimation in case 1 and 3.

Given that cases 2 and 3 exhibit similar evolution of rate constants, drug depletion dynamics, and concentration distribution within the DFPI, we assume a correlation between them. This leads us to hypothesize that regulating the convective-to-diffusive flux ratio, or equivalently $K/D_{eff}$, may enable precise tuning of rate constants. This control could be achieved by adjusting the porous media properties and effective diffusivity within the implant, allowing different regions to release the drug at distinct times—thus producing an increasing and decreasing rate constant to possibly simulate an on-and-off release behavior.

The parameters selected for this study, including Reynolds number, porosity, permeability, and effective diffusivity, are well within the range of values commonly reported in biomedical applications. For instance, Reynolds numbers in ocular applications are typically around 0.01 [18], while in the cardiovascular domain, they can reach up to approximately 2000 [22]. Similarly, effective diffusivity values in ocular systems have been reported to range from $10^{-9}$ to $10^{-14}$ m$^2$/s [24]. Permeability values, depending on the application, can be as low as $10^{-14}$ m$^2$ [25], highlighting the relevance of our parameter choices to realistic biomedical scenarios.

Our choice of porous media properties for DFPI that allowed us to achieve increasing rate constants at later stage (with prolonged depletion time) may have lessons for design of intelligent implants. It has been reported that, in addition to these porous media properties, release behavior can also depend on system dimensions [30]. While, we have not found any other study comparing the rate constant variation for cases with varying diffusive and convective fluxes, our results are consistent with literature reporting that release kinetics gets influenced by the porous media properties and system shape [31,32]. In the present work, we have assumed drug loading in DFPI, however, the drug loading arrangement is also reported to enable the control of release rate [33]. Laminated drug free layers can allow for achieving the zero-order release where its thickness allows for tuning of zero-order rate kinetics [34]. Implant geometry and fluid dynamics significantly influence drug release, with porous designs exhibiting more controlled profiles, particularly under varying flow conditions [25]. Adjusting implant shape to modify surface-to-volume ratios also impacts release kinetics [31]. DFPI's primary application lies in achieving sustained and predictable drug release profiles for various treatments, particularly in localized therapies where systemic exposure must be minimized. With the understanding developed in this work and other works in the literature capability of tuning of rate constants of porous implants will allow for precise control of drug delivery in biomedical applications.

## IV. CONCLUSIONS

This study emphasizes how porous media properties and flow conditions play a crucial role in shaping the drug release behavior of Drug-Filled Porous Implants (DFPIs). Through computational modeling, we explored the combined effects of permeability, porosity, effective diffusivity, and Reynolds number on drug release dynamics. The findings highlight how these factors influence the time it takes for the drug to deplete and the overall release pattern, providing valuable insights for improving the design and effectiveness of these implants.

1. For cases with lower permeability, a minimum drug depletion time from the implant is observed, regardless of variations in porosity and Reynolds number. In contrast, this lower limit diminishes and approaches zero for higher-permeability cases.

2. By fine-tuning Reynolds numbers, permeability, and effective diffusivity, the drug depletion rate can be slowed down, while also enabling a significant increase in the rate constant following an initial decline. This opens up exciting opportunities to design DFPI systems with controlled "on-and-off" release patterns, tailored to meet specific therapeutic goals.

3. Ratio of internal diffusive and convective fluxes in DFPI decide the depletion dynamics and drug distribution with time. It also appears to be correlated with the evolution of rate constants.

This study specifically focuses on the convection-diffusion-driven drug release from porous implants (DFPIs). However, other factors like erosion, swelling, pH changes, interactions with surrounding tissue, chemical reactions, and osmotic pressure could also play a major role, depending on the application and the material properties [15]. Additionally, the pulsatile flow often present in real-world scenarios might further influence the drug release behavior.

## ACKNOWLEDGMENTS

Authors thank the financial support of Anusandhan National Research Foundation, India, project number CRG/2021/001260. Authors also gratefully acknowledge the facilities provided by the Indian Institute of Technology Roorkee.

## ORCID

Pawan Kumar Pandey: 0000-0003-0145-3338
KVS Chaithanya: 0000-0003-0917-5619
Prateek K. Jha: 0000-0001-9844-2875